\documentclass[usenatbib,usegraphicx]{mn2e}
 
\usepackage{longtable}

\let\new=\newcommand
\new{\nodata}{...}
  
\title[Binary Star Astrometry \& Photometry from AEOS in 2002]{Astrometric and photometric measurements of binary stars with adaptive optics: observations from 2002}

\author[L.C. Roberts, Jr.]{Lewis C. Roberts, Jr.$^{1}$
\thanks{E-mail: lewis.c.roberts@jpl.nasa.gov}\\
$^{1}$ Jet Propulsion Laboratory, California Institute of Technology, 4800 Oak Grove Drive, Pasadena CA 91109, USA\\
}
\begin{document}

\date{}

\pagerange{\pageref{firstpage}--\pageref{lastpage}} \pubyear{2010}

\maketitle

\label{firstpage}

\begin{abstract}
 
The adaptive optics system at the 3.6 m AEOS telescope was used to measure the astrometry and differential magnitude in \textit{I}-band of 56 binary stars in 2002.  The astrometric measurements will be of use for future orbital determination, and the photometric measurements will be of use in estimating the spectral types of the component stars.    Two candidate companions were detected, but neither is likely to be gravitationally bound. Nine systems had not been observed in over 40 years. Eight of these are shown to share common proper motion, while HD 182352 is shown to be a background star. One of the two components of the HD 114378 ($\alpha$ Com) is shown to be a variable star of unknown type. In addition, 86 stars were unresolved and the full-width half maxima of the images are presented.

\end{abstract}



\textbf{Accepted to MNRAS}

\section{Introduction}

Speckle interferometry has been the dominant technique for observing visual binary stars for the last several decades. There are several groups actively monitoring binary stars with speckle interferometry.  One of the longest programs is that of the U.S. Naval Observatory \citep{mason2010}; others include the PISCO program \citep{prieur2010}, the WIYN speckle program \citep{horch2010} and that of Docobo and collaborators \citep{docobo2010}. The main purpose of these observations is measuring the astrometry of the system.  These measurements will eventually enable the computation of an orbit for the binary star, which leads to the determination of the dynamical masses of the component stars \citep{docobo2010}. To compute a high quality orbit, it requires many epochs of observations over at least one period of the orbit. For some orbits, this may take hundreds of years. 
 
Speckle interferometry systems are relatively inexpensive to build and straightforward to operate.  They are able to quickly observe large number of stars; as many as several hundred observations per night.  For all of its successes, the technique does have its  limitations.   It has a fairly low limiting dynamic range.  The exact limit  depends on the camera used, but ranges from 3-5 magnitudes \citep{mason1994}.   Also, in most cases true images are not created, but instead an autocorrelation, which can lead to quadrant ambiguity where the position of the companion is off by 180$^\circ$ \citep{bagnuolo1992}.  It is also difficult to extract the photometry from speckle interferograms, especially with the commonly used intensified detectors \citep{roberts1998}. 
  
Adaptive optics (AO) solves many  of the problems of speckle interferometry.  AO systems sense the phase aberrations in the incoming starlight and correct  it in real time.  Because the AO system corrects the atmospheric distortion, the images from the science camera have a much higher signal-to-noise ratio and can achieve a dynamic range of 10 magnitudes, though this depends on the separation between the star and the companion \citep{turner2008}.  With a coronagraph or image subtraction techniques the achievable dynamic range can be increased dramatically \citep{oppenheimer2009}. The other great advantage of AO is that it produces a true image of the system. This removes the quadrant ambiguity that is common in speckle interferometry, it also allows for the measurement of the photometry of the individual stars.  This information can be used to estimate the spectral type of the companion \citep{tenBrummelaar2000}.  This can then be compared to the mass (or mass sum) computed from the orbit.  If the observations are done in multiple filters, the components can be put on a colour-colour diagram \citep{caballero-nieves2010} for additional understanding of the system.  

Most adaptive optics observations have been focused on specific projects such as multiplicity surveys (e.g. Turner et al. 2008), but there is a great benefit to using AO for long term monitoring of binary stars.   The increased dynamic range of AO allows it to be used in the study of systems that the speckle interferometry has been unable to observe.  Many of these systems were discovered decades ago with visual methods (e.g. Burnham 1894) and the published astrometry has large errors.   

Between 2001 and 2006, the Advanced Electro-Optical System (AEOS) telescope and AO system \citep{roberts2002} were used to observe binary stars in \textit{I}-band in order to collect astrometric data to improve orbit determination and to provide photometric data for spectral class determination.  This paper presents the measurements from data collected in 2002.  The other observations will be be presented in subsequent papers. 
 
Most AO systems have science cameras that observe in the near-IR, only a handful of systems have operated in the visible. These include the Mt. Wilson system \citep{shelton1995} and the systems at the U.S. Air Force telescopes at the Starfire Optical Range \citep{fugate1994, spinhirne1998} and on Maui \citep{roberts2002}. During the last decade, only the U.S. Air Force's  AEOS telescope on Maui was available for astronomical observations and is currently unavailable for astronomical observations.  As such, photometric measurements from the AEOS telescope are unique and unlikely to be repeated in the near future. Photometric measuremetns in the visible are especially useful when combined with near-IR photometric measurements. The addition of visible measurements to near-IR measurements, decreases the uncertainty in spectral classification of stars.\citep{hinkley2010}. Of course, since the binary systems are dynamic, the astrometric measurements can not be duplicated.

\section{Observations}
 
Observations were made using the AEOS 3.6 m telescope and its AO system.  The AEOS telescope is located at the Maui Space Surveillance System at the summit of Haleakala \citep{bradley2006}. The AEOS AO system is a natural guide star system using a Shack-Hartmann wavefront sensor \citep{roberts2002}.   The individual subapertures have a diameter of 11.9 cm projected onto the primary.   The deformable mirror has 941 actuators.  The system's closed loop bandwidth is adjustable and can run up to 200 Hz, although the normal bandwidth is approximately 50 Hz.  In the configuration used for these observations, the light from  500-540 nm is sent to the tip/tilt detector system, the light from 540-700 nm is sent to the  wavefront sensor and the light longer than 700 nm is sent to the Visible Imager CCD science camera.

The observing list was created from the Washington Double Star Catalog (WDS) by selecting.  All objects had $V<8$ and $\delta>-25^\circ$.  The list included a number of binary stars with well measured astrometry, for comparison purposes. These were the stars with the smallest dynamic range in the observing list. It also includes stars with larger dynamic ranges than what speckle interferometry can do.  The intent was to gather additional astrometric measurements which could be used for eventual orbit calculation. Special attention was paid to binaries that had no recent published astrometry.  Many of these binaries have separations of several arcseconds, but their dynamic range is too large for  speckle interferometry observations to detect.  AO is well suited to observe these.   During testing and characterisation of the AO system a number of stars were observed that were not in the WDS.  I report on the results of these stars in addition to the observations of the known binaries. 

Each data set consists of 250 frames  using a Bessel \textit{I}-band filter. After collection, any saturated frames are discarded and the remaining frames are debiased, dark subtracted and flat fielded.  The frames are weighted by their peak pixel, which is proportional to their Strehl ratio and then co-added using a shift-and-add routine. The resulting image is analysed with the program FITSTARS; it uses an iterative blind-deconvolution that fits the location of delta functions and their relative intensity to the data.  The co-adding technique and the analysis with  FITSTARS was presented in \citet{tenBrummelaar1996} and \citet{tenBrummelaar2000}.  Observations were made in a queue scheduling mode and as such, observations were made during a wide range of observing conditions. 

Error bars on the astrometry and photometry were assigned using the method in \citet{roberts2005}. For the photometry, simulated binary stars were created from observations of single stars.  The photometry of these simulated binaries was measured and used to create a grid of measurement errors as a function of separation ($\rho$) and differential magnitudes.  For astrometry, the separation error bar is $\pm$0\farcs02 for $\rho$ $\leq$ 1\arcsec, $\pm$0\farcs01 for 1 $<$ $\rho$ $\leq$ 4\arcsec, and $\pm$0\farcs02 for $\rho$ $>$ 4\arcsec. The error in position angle ($\theta$) caused by errors in determining the centroid of the secondary star location will be larger for systems with small separations than wider separations.  I have adopted $\pm2^\circ$ for $\rho < 1$\arcsec and $\pm1^\circ$ for $\rho > 1$\arcsec as the position angle error. 

\section{Results}
 
The astrometry and photometry of all resolved systems are listed in Table \ref{binaries2}.  For each star, I list the Washington Double Star (WDS) number, the discovery designation, the HD Catalogue number, the Hipparcos Catalogue number, the Besselian date of the observation, the separation in arcseconds, the position angle in degrees and finally the differential magnitude measured in Bessel \textit{I}-band. Since the AEOS telescope is an alt-az design, it requires a Dove prism image derotator in the science camera to keep the orientation of the image fixed.  I carried out some tests of the derotator,  in which the derotator was turned off. In these cases, it was not possible to make an accurate measurement of the position angle of the stars.  The separation and differential magnitude are still published, while the separation will be of marginal value without a position angle, the differential magnitude is still valuable.   

The listed astrometry was compared with the latest published astrometry in the WDS. The astrometry for all but two systems was consistent with the published data. The position angle of  HD 137107 (WDS 15232+3017)  is inconsistent with the orbit of \citep{mason2006}, though the separation appears to be in agreement.   The position angle computed from the orbital elements is 77\degr85, while I measured 166\degr0. The observation is also inconsistent with published observations before and after the AEOS measurement \citep{bodin2003,horch2008}.  The position angle for HD 193215 (WDS 20154+6412) also does not match the orbit of \citep{seymour2002}, but again the separation is consistent.  The observations of Mason et al. (in preparation) collected in 2007 are consistent with the orbit of \citet{seymour2002}. It seems likely that in both cases the  image derotator malfunctioned.  
 
HD 114378 ($\alpha$ Com)  consists of an AB pair with an orbital semi-major axis of 0\farcs67 \citep{mason2006} along with a third star at a separation of over an arcminute.   Both of the measurements of the AB pair in this paper are off by 180\degr~from the astrometry computed from the orbit of \citet{mason2006}.  Looking at the data collected over the last few decades \citep{hartkopf2001}, quadrant ambiguity is a fairly common occurrence.  In addition, \citet{tenBrummelaar2000} measured the differential magnitude in \textit{I}-band as 0.0$\pm$0.03, while the images taken at AEOS, clearly show a pair of stars with unequal magnitude.  \citet{horch2010} also shows the differential magnitude varying from 0.0--0.66 over the course of a year. The 2002.2731 observation shown in Table \ref{binaries2} is the average of two measurements taken on the same night.  The measured differential magnitudes are  0.34, and 0.57. Some of the scatter is caused by relatively small separation of the pair, but the scatter is much higher than the scatter from other stars measured multiple times that same night.  The most likely explanation for all of these factors is that one of the components is a variable star with a relatively short period such as a $\delta$ Scuti or $\gamma$ Doradus type.

Nine of the stars  were last observed before 1970. The long time period is helpful in determining if the binaries share common proper motion. Table \ref{cpm} lists the  HD Catalogue number of the nine stars, the Besselian date, the change in position angle and separation from their last observation  \citep{mason2001} and finally the distance the secondary would have moved based on proper motion between the last observation and the one in this paper if it had no proper motion of its own.  The proper motion came from either the Hipparcos \citep{perryman1997} or Tycho catalogues \citep{hog1998}.  This shows that the companion to HD 182352 = WDS 19213+5543 is  a background star, while the other eight systems have common proper motion.

\subsection{Newly Discovered Components}
 
\begin{figure}
 \includegraphics[width=84mm]{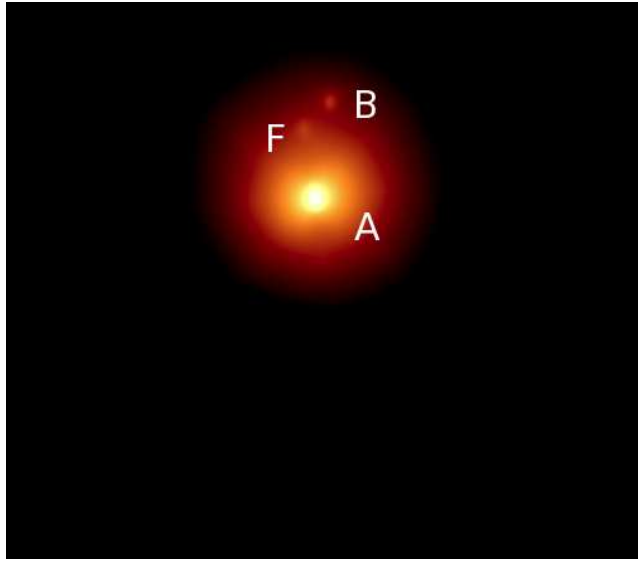}
 \caption{Image of HD 10425 observed on 2005.7727 showing the AB pair and the newly discovered F component.}
 \label{bu1103}
\end{figure}

\textit{HD 10425 (WDS 01433+6033)} This system is infrequently observed, probably due to its large dynamic range.   The central AB pair has only been observed three times since it was discovered in 1889 \citep{burnham1894} and it appears to have moved little since then.  As shown in Table \ref{cpm}, it is a common proper motion pair.    There are also C,D,E components with mean separations of approximately 1,2 and 3 arcminutes respectively.  A new component to this system was observed in 2002. An additional measurement in 2005 confirmed the component. The 2005 observation is shown in Fig. \ref{bu1103}.  The new component does not appear to form a hierarchical system with the A and B components, as the AB separation of 1\farcs6 is not 10 times the separation between the A component and the new component or the new component and the B component. As such it was tentatively labelled as F, but if it is found to be bound to the B component, it would be given the designation Bb. Follow up observations at a current epoch should be able to determine if the new component has the same proper motion as the AB pair.  

The primary has a spectral class of B8IIIn \citep{cowley1969}, which was certainly contaminated by the presence of the B component and possibly by the F component.  With only \textit{I}-band observations it is impossible to tell if either the B or F components fall on the main sequence or are also evolved stars.  

\textit{HD 118889 (WDS 13396+1045)}  This system already has three known components, a close pair with a separation of several arcseconds and a third component with a separation of several arcminutes. A new component was seen in two observations taken 61 days apart.  The position angle of the new component changed substantially between the two observations.  It does not appear to be an error in the image derotator, since the two observations of the AB pair are almost identical.    The object is either a field object with substantial proper motion or an artifact.  These have been seen before in AEOS data, but usually with a much higher dynamic range \citep{roberts2007}.  It is possible to estimate the companion spectral type assuming it was at the same distance as the primary and on the main sequence using the absolute magnitudes of the MK classification in \citet{cox2000}. Based on the primary's spectral class of a F0V \citep{royer2007}, and the differential magnitude in Table \ref{binaries2}, the companion would be an early M dwarf.  
                            
\subsection{Unresolved Stars}

Observations of stars that did not reveal a companion are detailed in Table \ref{single_stars2}.  The entry for each star lists the WDS number if it has one, the HD Catalogue number, the Hipparcos Catalogue number, the Besselian date of the observation, and the full width half maximum of the image (FWHM).  There are several reasons why known binary stars are not detected. The most common is that the known companion was within the central diffraction core of the image.  Second most common is that the companion was outside of the field of view. The Visible Imager camera has a 10$\arcsec$ field of view, but the star was usually not placed at the centre of the image due to a misalignment of the tip/tilt system and the AO system. The offset is usually about 1$\arcsec$.  Sometimes stars with separations of 4-5$\arcsec$ have their companion land outside of the field of view, eg. HD 159181 (WDS 17304+5218). 

These measurements are useful for several reasons.  For known binaries, they can constrain the orbital solution by providing an upper bound on the orbital separation at a specific time \citep{hinkley2010}. It also puts an upper limit to the brightness and separation of additional components of a system.  

HD 16295 (WDS 02366-1439)  has a single unconfirmed observation with a reported differential magnitude of 3.4 in the visible from the Hipparcos satellite \citep{perryman1997}. The image used in this paper has a fairly low signal to noise ratio due to poor AO correction and does not constrain the existence of the purported companion.  The observation only rules out additional bright companions. 

\section{Summary}

The AO system at the 3.6 m AEOS telescope was used to measure the astrometry and differential magnitude in \textit{I}-band of 56 binary stars in 2002. The astrometric measurements will be of use for future orbital determination, and the photometric measurements will be of use in estimating the spectral types of the component stars. In addition, 86 stars had no resolved companions,  and for these full-width half maxima of the images are presented.     

In addition, the observations allow for the analysis of specific stars.  Through analysis of differential magnitude from AO and speckle interferometry, it was determined that the  HD 114378 system contains a variable star with a relatively short period. Follow up measurements with CCD photometry are needed to determine the exact type of variable star. Simultaneous measurements with AO can be used to identify which component is the variable. The proper motion of nine stars that had not been observed in decades were analysed. HD 182352 was determined to be a background star, while the rest share common proper motion.

Two possible new companions were detected to already known binary stars.  A candidate companion to HD 10425 was detected and confirmed with follow up observations several years later. While the system does not appear hierarchal, it is possible it is a physical system with a high inclination. Follow up multi-filter near-IR AO observations  are needed to determine if it is a physical system through analysis of proper motion and  colour photometry.  The candidate companion to HD 118889 has a smaller probability of being physical, but follow up observations are needed to determine if the companion is physical or an artifact.


\section{Acknowledgements}

I thank Kimberly Nguyen for years of love and support.  I thank Brian Mason and Bill Hartkopf for providing data from the WDS Catalogue and useful and insightful comments.  Also I thank the numerous staff members of the Maui Space Surveillance System who helped make this data possible.  The research in this paper was carried out at the Jet Propulsion Laboratory, California Institute of Technology, under a contract with the National Aeronautics and Space Administration. Additional funding came from AFRL/DE (Contract Number F29601-00-D-0204).  This research made use of the Washington Double Star Catalogue maintained at the U.S. Naval Observatory, the SIMBAD database, operated by the CDS in Strasbourg, France and NASA's Astrophysics Data System.



\onecolumn

\begin{longtable}{llrrcrrc} 
  \caption{Resolved Binaries}\label{binaries2}\\

   \hline 
    WDS & Discovery & HD & HIP & Epoch & $\rho$ & $\theta$&  $\Delta$ \textit{I}\\
    \#  & Designation & \# & \#  &       &(\arcsec)& (\degr) & \\
    \hline
\endhead

\hline
\multicolumn{2}{l}{Continued on next page}
\endfoot

\hline 
\endlastfoot

00554$+$4023 & A 1511     & 5314    &  4331  & 2002.7473 &  1.21 &  39.6  &            2.39$\pm$0.01\\
01060$-$0840 & RST 4165   & 6514    &  5154  & 2002.6244  &  1.24  &   44.9    &       3.92$\pm$0.04\\
01274$+$0658 & A 2006 AB     & 8849    &  6788  & 2002.6245 &  2.02 &  276.3 &          4.39$\pm$0.01\\
01277$+$4524 & BU 999 AB     & 8799    &  6813  & 2002.6052  &  0.70  &  325.2    &     4.7$\pm$0.2\\
01315$+$1521 & BU 506     & 9270    &  7097  & 2002.6052  &  0.61  &  292.6    &       4.46$\pm$0.15\\
01418$+$4237 & MCY 2      & 10105   & 7918   & 2002.7773  &  0.75  &  212.0    &       5.5$\pm$0.4\\
01433$+$6033 & BU 1103 AB       & 10425  & 8046 & 2002.7773  &  1.61  &  358.2    &          4.71$\pm$0.09\\
      ''     &     ''       &    ''     &   ''     & 2005.7727  &  1.62  &  359.6    &       3.8$\pm$0.1\\
01433$+$6033 & New  &  10425  & 8046 & 2002.7773  &  1.15  &  340.8    &         4.99$\pm$0.11\\
      ''     &     ''       &    ''     &   ''     & 2005.7727  &  1.18  &  342.2    &       4.17$\pm$0.06\\ 
02291$+$6724 & CHR 6 Aa,Ab & 15089   & -& 2002.6053  &  0.44  &  238.9    &       2.86$\pm$0.04\\ 
02291$+$6724 & STF 262 AB  & 15089   &- & 2002.6053  &  2.70  &   44.0    &       2.25$\pm$0.03\\  
02449$+$1007 & TOK 1 Aa,Ab & 17094   &12828   & 2002.7804  &  0.21  &  213.5    &       3.39$\pm$0.20\\ 
03027$-$0741 & BU 11      & 18953   &14168   & 2002.7804  &  1.37  &   63.3    &       3.93$\pm$0.04\\
03524$+$1836 & HDS483     & 24278   & 18116  & 2002.9743  &  0.63  &  -  &       3.25$\pm$0.02\\  
04134$+$6556 &  HDS536    & 26221   &  19710 & 2002.9743  &  0.42  &  -  &       4.77$\pm$0.28\\
05182$+$3322 & STT 103    & 34334   &  24727 & 2002.9881  &  4.14  &   55.3    &       6.70$\pm$0.17\\
05282$-$2046 & BU 320 AB  & 36079   &  25606 & 2002.7202  &  2.58  &    1.4    &        5.20$\pm$0.04\\
05597$+$3713 & STT 545 AB & 40312   &  28380 & 2002.7529  &  3.91  &  304.9    &        4.53$\pm$0.02\\
06532$+$3826 & COU 1877   & 50037      &  33064 & 2002.8023  &  0.27  &  187.1    &    1.93$\pm$0.03\\
07176$+$0918 & STT 170    & 56515      &  35310 & 2002.2731  &  0.35  &   48.1    &    0.15$\pm$0.02\\   
07518$-$1354 & BU 101     & 64096      &  38382 & 2002.2731  &  0.14  &   15.8    &    0.70$\pm$0.03\\%
08122$+$1739 & STF 1196 AB& 68257      &  40167 & 2002.2731  &  0.88  &   70.35   &     0.26$\pm$0.03\\ 
08468$+$0625 & SP 1AB     & 74874      &  43109 & 2002.2731  &  0.24  &  228.8    &    1.88$\pm$0.03\\ 
      ''     &     ''       &    ''     &   ''     & 2002.2732  &  0.24  &  226.5    &    1.91$\pm$0.03\\ 
08468$+$0625 & STF 1273 AB-C&74874 &   43109 & 2002.2731  &  2.85  &  293.9    &        3.87$\pm$0.01\\
      ''     &     ''       &    ''     &   ''      & 2002.2732  &  2.84  &  294.5    &       3.59$\pm$0.01\\ 
09006$+$4147 & KUI 37 AB   & 76943  &  44248  & 2002.2731  &  0.71  &   20.5    &      1.89$\pm$0.03\\ 
      ''     &     ''       &    ''     &   ''     & 2002.2732  &  0.70  &   20.3    &       1.89$\pm$0.03\\ 
      ''     &     ''       &    ''     &   ''     & 2002.2733  &  0.70  &   20.1    &       1.87$\pm$0.03\\ 
09154$+$2248 & A 2136     &   79553  &  45425& 2002.1477 & 1.85 & 111.8  &             3.8$\pm$0.01 \\
09167$-$0621 & KUI 40     &79910  &   45527  & 2002.1476 & 1.44 & 283.6              & 6.0$\pm$ 0.2  \\   
11182$+$3132 & STF 1523 AB&  98230 &  - & 2002.0192  &  1.75  &  259.7    &       0.55$\pm$0.01\\
      ''     &     ''       &    ''     &   ''      & 2002.2731  &  1.78  &  259.4    &       0.61$\pm$0.01\\
11239$+$1032 & STF 1536 AB &  99028  &   55642& 2002.1697 &1.76    & 106.5    &         2.27$\pm$0.01   \\   
12244$+$2535 & STF 1639 AB& 108007 &   60525 & 2002.1068  &  1.72  &  332.3    &        0.92$\pm$0.01\\   
      ''     &     ''       &    ''     &   ''     & 2002.2731  &  1.72  &  322.7    &       0.92$\pm$0.01\\ 
12540$+$5558 & BLM   2    & 112185 &   62956 & 2002.2732  &  0.11  &   59.9    &       2.31$\pm$0.03\\
13099$-$0532 & MCA 38 Aa,Ab& 114330  &   64238 & 2002.4702& 0.44  & 340.0   &          2.21$\pm$0.03   \\ 
13100$+$1732 & STF 1728 AB & 114378 &   64241 & 2002.0988  &  0.19  &  194.4    &       0.210$\pm$0.04\\ 
      ''     &     ''       &    ''     &   ''     & 2002.2731  &  0.20  &  190.3    &       0.49$\pm$0.04\\ 
13396$+$1045 & BU  612 AB  & 118889 &   66640 & 2002.1069  &  0.20  &  179.2    &       0.20$\pm$0.04\\    
      ''     &     ''       &    ''     &   ''     & 2002.2732  &  0.21  &  179.5    &       0.18$\pm$0.04\\
13396$+$1045 & New       & 118889 &   66640& 2002.1069  &  3.59  &  135.9 &         5.40$\pm$0.06\\  
      ''     &     ''       &    ''     &   ''     & 2002.2732  &  3.61  &   76.2      &       5.9$\pm$0.06\\ 
14118$-$1901 & DON 649    &  124087  & 69356 & 2002.4758  &  2.19  & 303.2     &       4.9$\pm$0.03\\  
14148$+$1006 & KUI 66     & 124679  &  69612 & 2002.3502  &  0.98  & 106.1     &       4.2$\pm$0.06  \\             
14171$-$1835 & HDS 2008   & 124990  &   69792& 2002.3503  & 0.67 & 88.5 &              5.4$\pm$0.45 \\     
14227$+$0216 & HDS 2025   & 143275 &  78401  & 2002.5578  & 0.73 &  3.1     &          2.8$\pm$0.03\\
15232$+$3017 & STF 1937 AB & 137107 &   75312 & 2002.1342  &  0.61  &   166.0   &       0.29$\pm$0.03\\
15576$+$2653 & AGC   7 AB  & 143107 &   78159 & 2002.5797  &  1.96  &     9.8   &       6.44$\pm$0.13\\ 
16304$+$4044 & HDS 2331   & 149025 &  80827  & 2002.4706  &  0.56  &275.2    &         3.32$\pm$0.07  \\
16492$+$4559 & BU  627 AB    & 152107 &   82321 & 2002.5553  &  1.79  &  33.4  &        3.92$\pm$0.04\\
16492$+$4559 & BU  627 AC    & 152107 &   82321 & 2002.5553  &  1.95  &  31.7  &        3.33$\pm$0.04 \\  
16492$+$4559 & A 1866 BC    & 152107 &   82321 & 2002.5553  &  0.15  &  197.1 &        0.53$\pm$0.04 \\
18000$+$2449 & COU 115    & -&  -& 2002.5473  &  0.28  &   113.4   &       0.53$\pm$0.04\\
18272$+$0012 & STF 2316 AB    & 169985 &   90441 & 2002.5611  &  3.70  &  318.6   &      3.42$\pm$0.01 \\
18582$+$1722 & HO 91 AB   & 176155 &   93124 & 2002.5611  &  6.79  &   143.7   &       6.33$\pm$0.11\\
19122$+$3215 & HU  941    & 179708 &   94350 & 2002.5803  &  1.07  &   145.2   &       2.890$\pm$0.01 \\
19177$+$2302 & BU  248 AB    & 180968 &   94827 & 2002.5803  &  1.72  &   127.2   &     3.57$\pm$0.03\\
19213$+$5543 & A  1395    & 182352 & -  & 2002.5804  &  4.02  &   233.9   &       2.00$\pm$0.01\\
19244$+$1656 & HDS 2753 Aa,Ab & 182490 &   95398 & 2002.5802  &  0.43  &   138.1   &     3.44$\pm$0.08\\
19364$+$5013 & BU 1131 AB    & 185395 &   96441 & 2002.6920  &  2.54  &    65.8   &     5.89$\pm$0.09\\
19556$+$5226 & YR 2 Aa,Ab    & 189037 &   98055 & 2002.5805  &  0.10  &    77.6   &     0.66$\pm$0.05 \\
19556$+$5226 & STF 2605 AB  & 189037 &   98055   & 2002.5805  &  2.88  &   174.1   &     2.46$\pm$0.01 \\
20154$+$6412 & MLR  60 AB    & 193215   &   99832 & 2002.5831  &  0.19  &   184.1   &   0.12$\pm$0.04\\
20176$-$1230 & WZ 15 Aa,Ab &192876 & 100027    & 2002.7277  & 0.82  & 352.6  &         6.4$\pm$0.6 \\
20569$-$0942 & BU 1034 AB  & 199345   &  103401 & 2002.5857  &  2.12  &   164.5   &     6.15$\pm$0.11\\
21103$+$1008 & KNT   5 AB  & 201601   &  104521 & 2002.5857  &  0.96  &   258.0   &     3.82$\pm$0.04\\
22051$+$5142 & HDS 3134    & 209870   &  109015 & 2002.7470  &  0.98  &   352.9   &     3.05$\pm$0.01\\
22332$+$3356 & HO 293     &213745	   & - & 2002.6352  & 1.51 & 314.9       &     2.35$\pm$0.01 \\
23300$+$5833 & STT 496 AB  & 221253   &  115990 & 2002.7116  &  0.78  &   355.9   &     3.66$\pm$0.05\\
23355$-$0709 & RST 4726    & 221823   &  116426 & 2002.7117  &  1.07  &   277.0   &     3.71$\pm$0.03\\
\end{longtable}

\newpage
 
\begin{table}
  \caption{Proper Motion of Neglected Binaries}\label{cpm}
  \begin{tabular}{rlccc}
  \hline
HD  & Date     &    $\Delta \rho$  &  $\Delta \theta$  &    PM Movement\\
\#   & Last Obs &      (\arcsec)    &    (\degr)        &      (\arcsec)\\
 \hline
8849    & 1953.03     & 0.44     &   3.3       &  6.8\\
8799    & 1967.387  &   1.07     & 206.8        & 13.45 \\
10425   & 1934.70   &   0.38      & 16.1      & 1.48 \\
34334   & 1936.21   &   0.27      &  0.9      &10.9  \\
79553   & 1933.30   &   0.07      & 10.9      & 1.35 \\
143107  & 1965.251  &   0.14    &    0.6     &     3.64\\
182352  & 1929.57   &   3.24    &   19.6    &     2.91\\
185395  & 1968.723  &      0.38   &     5.9  &   8.9\\
\hline
\end{tabular}
\end{table}

\begin{longtable}{lrrcl}
 \caption{Unresolved Stars} \label{single_stars2}\\
  \hline
   WDS  & HD & HIP & Epoch & FWHM\\
   \#   & \# & \#  &       &(\arcsec) \\
 \hline
\endhead

\hline
\multicolumn{2}{l}{Continued on next page}
\endfoot

\hline
\endlastfoot

00542$+$4318  &    5178  &  -  & 2002.7473 & 0.26     \\ 
01100$+$5202  &    6843  & 5468  & 2002.6245 & 0.13     \\    
01334$+$5820  &    9352  &    7251         & 2002.7773 & 0.14    \\      
02095$+$3459  &   13161  &   10064          & 2002.6053 & 0.13      \\ 
02130$+$0851  &   13611  &   10324          & 2002.6053 & 0.16      \\   
02132$+$4414  &   13520  &   10340         & 2002.6053 & 0.14      \\      
02136$+$5104  &   13530  &   10366          & 2002.7773 & 0.19     \\      
02171$+$3413  &   13974  &   10644          & 2002.7773 & 0.13     \\     
02366$-$1439  &   16295  &   12146        & 2002.7475 & 0.38     \\  
02366$+$1227  &   16234  &   12153         & 2002.7475 & 0.16     \\                   
02424$+$2001  &   16811  &   12640         & 2002.7804 & 0.12      \\        
02422$+$4012  &   16739  &   12623          & 2002.7529 & 0.13     \\                
02500$+$2716  &   17573  &   13209         & 2002.7503 & 0.24      \\                 
02543$+$5246  &   17879  &   13531         & 2002.7503 & 0.21      \\                
03429$+$4747  &   22928  &   17358         & 2002.7504 & 0.18      \\    
03460$+$6321  &   23089  &   17587         & 2002.6898 & 0.11      \\                 
03492$+$2403  &   23850  &   17847         & 2002.7504 & 0.18      \\                
03501$+$4458  &   23838  &   17932         & 2002.9743 & 0.10     \\               
04230$+$1732  &   27697  &   20455          & 2002.6898 & 0.13    \\    
    -        &   32630  &   23767         & 2002.6545 & 0.10     \\                  
     ''       &    ''      &   ''              & 2002.7147 & 0.10     \\  
05154$+$3241  &   33959  &   24504          & 2002.7257 & 0.11  \\  
05251$+$0621  &   35468  &   25336    & 2002.7393 & 0.24        \\  
05263$+$2836  &  35497   & 25428   & 2002.7448 & 0.09           \\  
      -      &   39040  &   27383& 2002.0158 & 0.39     \\   
05490$+$2434  &  38751   &27468     & 2002.7145 & 0.16    \\     
     -       &  39190   &27533 & 2002.7721 & 0.35     \\   
     -       & 168434   &  - & 2002.7146 & 0.08     \\                  
06230$+$2231  &   44478  &   30343         & 2002.7529 & 0.11       \\
     -       &   48737  &   32362       & 2002.7530 & 0.13      \\                
07269$+$2015  &   58579  &   36156         & 2002.3334 & 0.24      \\                
08447$+$1809  &   74442  &   42911          & 2002.3334 & 0.26      \\               
     -       &   86535  &   48935    & 2002.1477 & 0.14     \\                 
11239$+$1032  &   99028  &   55642       & 2002.1697 & 0.17       \\  
     -       &  102047  &   57264       & 2002.1697 & 0.41     \\           
     -       &  105707  &   59316   & 2002.0904 & 0.11     \\               
13022$+$1058  &  113226  &   63608   & 2002.1069 & 0.14           \\ 
13189$-$2310  &  115659  &   64962    & 2002.0987 & 0.17       \\ 
      -      &  123123  &   68895       & 2002.1069 & 0.12     \\        
14141$+$1258  &  124570  &   69536         & 2002.3502 & 0.14      \\          
14157$+$1911  &  124897  &   69673       & 2002.4757 & 0.18      \\     
15245$+$3723  &  137391  &   75411        & 2002.4022 & 0.18      \\                
15249$+$5858  &  137759  &   75458  & 2002.2055 & 0.10          \\ 
     -       &  139006   &     76267 & 2002.2732 & 0.12     \\                  
16003$-$2237  &  143275  &   78401   & 2002.1262 & 0.14        \\  
16143$-$0342  &  146051  &   79593  & 2002.4842 & 0.14          \\   
     ''       &    ''      &    ''      & 2002.0987 & 0.08     \\                 
16183$-$0442  &  146791  &   79882      & 2002.1262 & 0.12     \\  
16221$+$3054  &  147677  &   80181         & 2002.5798 & 0.14      \\                
16254$+$3724  &  148283  &   80460        & 2002.4706 & 0.08      \\               
-    &  153210   &   -   83000    & 2002.3176 & 0.13     \\  
17003$+$3056  &  153808  &   83207         & 2002.5553 & 0.10      \\                 
17150$+$2450  &  156164  &   84379          & 2002.5117 & 0.20  \\  
      ''      &     ''     &        ''          & 2002.5552 & 0.12     \\                 
17304$+$5218  &  159181  &   85670      & 2002.3067 & 0.16     \\  
17395$+$4600  &  160762  &   86414          & 2002.5612 & 0.12    \\                 
-    &  161096  &   86742& 2002.3177 & 0.12     \\                 
17566$+$5129  &  163967  &   87833      & 2002.3203 & 0.09      \\              
-    &  163917  &   88048 & 2002.3204 & 0.09     \\  
-     &  168411  &   89140  & 2002.5446 & 0.41 \\h
18210$-$2950  &  168454  &   89931    & 2002.4352 & 0.18      \\   
18213$-$0254  &  168723  &   89962    &  2002.4352 & 0.14      \\ 
18280$-$2525  &  169916  &   90496     & 2002.3998 & 0.12     \\    
18280$+$0612  &  170200  &   90497        & 2002.5611 & 0.17      \\                
18367$+$0640  &  171834  &   91237         & 2002.5611 & 0.12     \\                
18426$+$5532  &  173524  &   91755          & 2002.5611 & 0.12     \\             
18553$-$2618  &  175191  &   92855      & 2002.3999 & 0.17      \\  
18589$+$3241  &  176437  &   93194    & 2002.3999 & 0.11       \\              
19054$+$1352  &  177724  &   93747      & 2002.4518 & 0.12      \\            
19126$+$6740  &  180711  &   94376    & 2002.4847 & 0.19       \\    
19164$+$1433  &  180555  &   94720         & 2002.5803 & 0.14    \\                
19180$+$2012  &  181025  &   94847        & 2002.5804 & 0.25      \\             
19190$+$3727  &  181470  &   94932          & 2002.5803 & 0.19      \\            
19205$-$0525  &  181391  &   95066         & 2002.5802 & 0.13      \\
19463$+$1037  &  186791  &   97278    & 2002.5503 & 0.10         \\ 
19508$+$0852  &  187642  &   97649    & 2002.5556 & 0.12        \\   
19542$+$0828  &  188310  &   97938        & 2002.5529 & 0.19     \\             
20143$+$0803  &  192366  &   99740       & 2002.5830 & 0.37     \\         
20158$+$2749  &  192806  &   99874      & 2002.5831 & 0.11      \\                
20222$+$4015  &  194093  &  100453    & 2002.5664 & 0.18     \\   
20349$+$2833  &  196198    &  -   &  2002.5777  & 0.12   \\
20414$+$4517  &  197345  &  102098     & 2002.5637 & 0.11      \\               
20527$-$0859  &  198743  &  103045         & 2002.5857 & 0.12   \\               
21186$+$6235  &  203280  &  105199    & 2002.5665 & 0.16     \\      
21567$+$6338  &  208816  &  108317         & 2002.7058 & 0.12    \\                
21573$+$5029  &  208785  &  108374       & 2002.7058 & 0.13      \\             
22038$+$6438  &  209790  &  108917         & 2002.7470 & 0.14      \\                
22236$+$5214  &  212496  &  110538        & 2002.5586 & 0.11     \\                
22350$+$6050  &  214165  &  111469      & 2002.6352 & 0.25      \\                 
\end{longtable}

\label{lastpage}

\end{document}